\documentclass{iau} 
\usepackage{graphicx}
\usepackage{color}
\title[Accretion Rate in AGNs] 
{Accretion rate in AGN and X-ray-to-optical flux ratio at z $\le$ 0.2}

\author[Gaulle et. al.]   
{Asrate Gaulle$^{1,2}$,
	Mirjana Povi\'c$^{2,3}$,
	and Dejene Zewdie$^{4,5}$ }

\affiliation{$^{1}$ Dilla University, Department of Physics, Dilla, Ethiopia; \textbf{email: {asrieguale@gmail.com}}
	$^{2}$Astronomy and Astrophysics Research and Development Division, Ethiopian Space Science and Technology Institute, Addis Ababa, Ethiopia; $^{3}$Institute of Astrophysics of Andaluc\'ia (IAA-CISC), Department of Extragalactic Astronomy, Granada, Spain;
	$^4$ N\'ucleo de Astronom\'ia, Universidad Diego Portales, Santiago, Chile;
	$^{5}$ Department of Physics, Debre Berhan University, Debre Berhan, Ethiopia
 }

\pubyear{2019}
\volume{356}  
\jname{Nuclear Activity in Galaxies Across Cosmic Time}
\editors{M. Povi\'c, P. Marziani, J. Masegosa, H. Netzer,\\ S. H. Negu, \&
	S. B. Tessema,  eds.}
\begin{document}
	
	\maketitle
	
	\begin{abstract}
		We explored a sample of 545 local galaxies using data from the 3XMM-DR7 and SDSS-DR8 surveys. We carried out all analyses up to z\,$\sim$\,0.2, and we studied the relation between X/O flux ratio and accretion rate for different classes of active galaxies such as LINERs and Seyfert 2. We obtained a slight correlation between the two parameters if the whole sample of AGN is used. However,
		LINERs and Sy2 galaxies show different properties, slight correlation and slight
		anti-correlation, respectively. This could confirm that LINERs and Sy2 galaxies
		have different accretion efficiencies and maybe different accretion disc properties,
		as has been suggested previously.
		
		\keywords{galaxies - active; AGN - accretion rate; AGN - black hole masses; AGN - X-ray properties; AGN - optical properties.}
	\end{abstract}
	
	\firstsection 
	\section{Introduction}
	Active galactic nuclei (AGN) are powerful sources of radiation in a wide spectral range, from gamma-rays to radio waves (Netzer 2015). In particular, AGN are strong X-ray sources. X-ray emission is shown to be a powerful tool of AGN detection and a study of the growth of supermassive black holes (SMBHs) and AGN properties (Brandt and Hasinger 2005). On the other side, optical data are very important for AGN classification and for studying the properties of AGN host galaxies (Povi\'c et al. 2009a,b, 2012). Therefore, the combination of X-ray and optical data allows the successful study of the connection between the AGN and their host galaxies. Furthermore, the accretion rate (AR) in galaxies remains a prerequisite for understanding the physics behind SMBHs and AGN, the evolution and growth of galaxies, and the connection between active and non-active galaxies. However, AR measurements are still not easy since they mainly depend on the availability of spectroscopic data, which contain smaller data sets and poorer statistics than photometric data. In previous studies, Povi\'c et al. (2009a, b) suggested that there might be a photometric indicator of AR in galaxies, based on the ratio between the flux in X-rays (0.5 - 4.5 keV) and optical flux. In this work, we used X-ray data, and optical spectroscopic and photometric data to measure both X/O flux ratio and AR in different types of active galaxies and to test the correlation between the two parameters. 
	\section{Data and Sample Selection}
	X-ray data were obtained from the 3XMM-DR7 catalogue (Rosen et al. 2016) which contains 727,790 sources. Also, we used optical photometric and spectroscopic data from the Sloan Digital Sky Survey (SDSS) data release 8 (DR8). Spectroscopic measurements have been obtained from the MPA-JHU SDSS DR8 catalogue (Brinchmann et al. 2004) for 1,472,581 sources. Finally, morphological classification has been obtained from the Galaxy Zoo catalogue (Lintott et al. 2011). We cross-matched all three catalogues using a cross-matched radius of 3 arcsec and we selected 2151 sources in total up to redshift z\,$\le$\,0.2. We stick to this redshift for avoiding the K-correction in X-rays. Using the BPT-NII diagram (Baldwin, Phillips, \& Terlevich, 1981), and signal-to-noise ratio of emission lines of S/N\,$>$\,3, we selected 545 AGN galaxies in total, of those 209 and 336 being Seyfert 2 and LINER sources, respectively. 
	
	\section{Analysis and Results}
	In this work, we analyzed the correlation between the X/O flux ratio and AR of spectroscopically selected AGN up to z\,$\le$\,0.2. After selecting LINERs and Sy2 sources, we went through the following analysis: we first measured the X/O flux ratio by using the (0.5 - 4.5keV) X-ray and optical r bands (Povi\'c et al. 2009a, b). Secondly, we measured the mass of a black hole using the velocity dispersion method (Tremaine et al. 2002) and velocity dispersion measurements from the MPA-JHU catalogue (Brinchmann et al. 2004). Eddington luminosity was then measured as L$_{edd}$\,=\,1.5\,$\times$\,10$^{38}$ (Mbh/M)erg/s. Bolometric luminosity was measured through the H$\beta $ and [OIII] emission lines using the results of Netzer (2013). Emission lines were first corrected for extinction through the H$\alpha$/H$\beta$ emission lines. Finally, the accretion rate was measured as AR = Lbol /Ledd. We tested the correlation between the SMBH mass and AR (see Fig. 1) and X/O flux ratio and AR (see Fig. 2) for the whole sample of AGN, and also for LINERs and Sy2 galaxies. We finally analyzed the same relations, but for different morphological types.  We found anti-correlation between the SMBH mass and AR independently on AGN type, as shown in Fig. 1, confirming some of the previous results (e.g., Woo and Urry 2002). Only mild correlation has been found between the X/O flux ratio and AR in LINERs, and mild correlation in Seyfert 2 galaxies, as shown in Fig. 2.  between X/O flux ratio and AR as shown inFig. 3. When observing the previous in relation to morphology, the same trends have been found in all cases for both elliptical and spiral galaxies at redshifts z\,$\le$\,0.2. 
	
	\begin{figure}[!h]
		\centering
		\includegraphics[width=6.6cm, height=5.465cm]{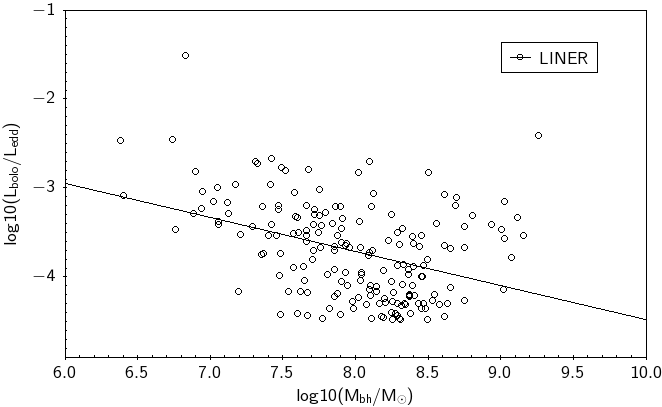}
		\includegraphics[width=6.6cm, height=5.465cm]{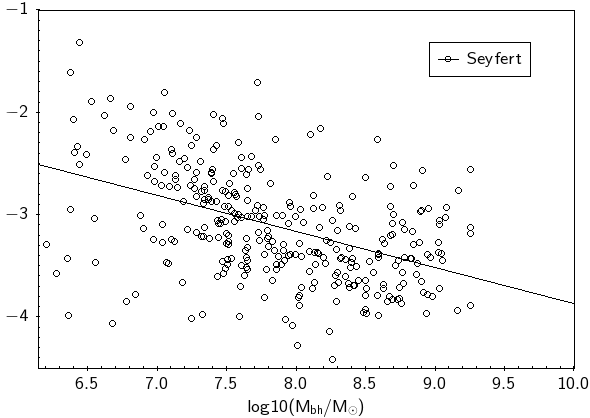}
		\caption{The correlation between BH mass and AR, left-right (LINER,Sy2)}
		\label{fig:1}
	\end{figure}
	
	\begin{figure}[!h]
		\centering
		\includegraphics[width=6.6cm, height=5.465cm]{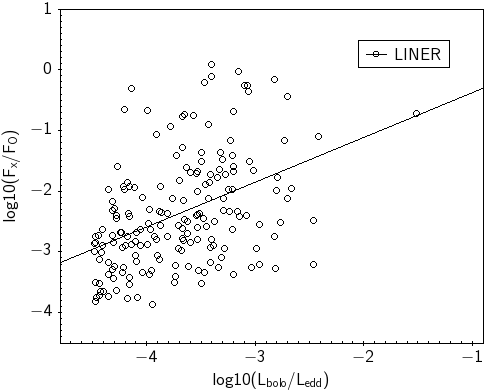}
		\includegraphics[width=6.6cm, height=5.465cm]{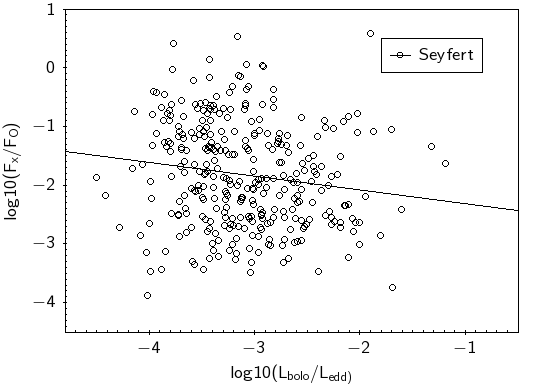}
		\caption{The correlation between  X/O flux ratio \& AR, left-
			right: LINER, Sy2.}
		\label{fig:2}
	\end{figure}
	\section{Conclusion}
	The results obtained above could suggest previous findings that Sy2 and LINER galaxies have different accretion
	properties and belong to two different modes: radiatively efficient
	and radiatively inefficient advection dominated accretion. Preliminary results obtained in this work do not discard the possibility of using the X/O flux ratio as an AR indicator, however, more detailed studies and larger statistical samples are needed for confirming this. 
	
	\section*{Acknowledgments}    
	AG acknowledges the support from Dilla University. AG and MP acknowledge financial support from the EORC under ESSTI.
	Mirjana Povic acknowledges the support from the Spanish Ministry of Science, Innovation, and Universities 
	(MICIU) through project AYA2016-76682C3-1-P.

	\section*{Reference}
	
	\noindent
	Baldwin, J. A., Phillips, M. M. \& Terlevich, R., 1981, PASP, 93, 5\\
	Brandt, W. N., \& Hasinger, G., 2005, ARA\&A, 43, 827\\
	Brinchmann J., Charlot S., White S. D. M., et al., 2004, MNRAS, 351, 1151\\
    Lintott, C., et al., 2011, \textit{MNRAS}, 410, 166\\ 
	Netzer H., 2013, Cambridge University Press\\
    Netzer H., 2015, ARA\&A, 53, 365\\
	Povi\'c, M., et al., 2009a, \textit{ApJ}, 702, 51\\     
	Povi\'c, M., et al., 2009b, \textit{ApJ}, 706, 810\\
    Pov\'ic, M., S{\'a}nchez-Portal, M., and Perez-Garc{\'\i}a, A., et al., 2012, A\&A, 541, 118\\
	Rosen, S. R., et al., 2016, A\&A, 590, 1\\
    Tremaine, S., et al. 2002, ApJ, 574, 740\\
	Woo, J.-H., \& Urry, C. M., 2002, \textit{ApJ}, 579, 530\\
	%
	
\end{document}